\documentstyle[twocolumn,aps,pre]{revtex}
\newcommand{\g}[1]{{\bf {#1}}}
\begin{document}

\title{Slow inviscid flows of a compressible fluid in spatially
inhomogeneous systems}
\author{V.P. Ruban \cite{1}}
\address {\it Optics and Fluid Dynamics Department, 
Ris\o ~National Laboratory, DK-4000 Roskilde Denmark \\
and  
L.D.Landau Institute for Theoretical Physics,
2 Kosygin Street, 117334 Moscow, Russia}
\date{June 2001}
\maketitle

\begin{abstract}
An ideal compressible fluid is considered, with an equilibrium density
being a given function of coordinates due to presence of some static
external forces. The slow flows in such system, which do not disturb the 
density, are investigated with the help of the Hamiltonian formalism. 
The equations of motion of the system are derived for an arbitrary given
topology of the vorticity field. The general form of the Lagrangian for 
frozen-in vortex lines is established. The local induction approximation 
for motion of slender vortex filaments in several  
inhomogeneous physical models is studied.
\end{abstract}

\medskip



\section{Introduction}

Hydrodynamic-type systems of equations are extensively employed for
macroscopic description of physical phenomena in ordinary and superfluid
liquids, in gases, in plasmas, in other substances. In solving
hydrodynamic problems, it is admissible in many cases to neglect all
dissipative processes and use the ideal fluid approximation, at least as
the first step. With this approximation, a dynamic model describing flow
is conservative. The Hamiltonian formalism is a convenient tool to deal
with such systems \cite{Arnold},\cite{DNF}, 
which makes possible to consider in a universal way all
nonlinear processes. A big number of works is devoted to application
of the Hamiltonian method in hydrodynamics (see, for instance, the reviews
\cite{ZK97},\cite{M98} and references therein).

One of the most important questions, permitted for a universal consideration in
the frame of canonical formalism, is the question about integrals of
motion of a dynamic system. Accordingly to the theorem of Noether
\cite{Arnold},\cite{DNF}, each conservation law of a system is closely 
connected with a symmetry of the corresponding
Lagrangian with respect to some one-parametric group of transformations of
dynamical variables. It is well known that the conservation laws for
the energy, momentum, and for the angular momentum follow from the
fundamental properties of the space and time, namely from homogeneity of
the time and from homogeneity and isotropy of the space. Due to these
properties, shifts and rotations of a system do not change its 
Lagrangian. The characteristic 
feature of the hydrodynamic-type systems is that they possess, besides the
indicated usual integrals of motion, also an infinite number of specific 
integrals of motion related to the freezing-in property of canonical 
vorticity \cite{ZK97}-\cite{KR2000}. The reason for this is a basic 
physical property of fluids, relabeling symmetry. For instance, in isentropic
flows the circulation of the canonical momentum along any frozen-in closed
contour is conserved. In usual nonrelativistic hydrodynamics, where the
canonical momentum coincides with the velocity, the given statement is known
as the theorem of Kelvin about conservation of the velocity circulation
\cite{Lamb},\cite{LL6}. 

Existence of an infinite number of integrals of motion influences
strongly dynamical and statistical properties of liquid systems. 
This is the reason why a clarification of structure  of conservation
laws is very important, as well as search for such new parameterizations 
for dynamical variables, which take into account the integrals of motion 
more completely. In many cases, even when a dissipation is present but
its level is low, it is still correct to speak about integrals of the
corresponding conservative problem, because values of some of them are
conserved with a high accuracy, especially on an initial stage of the
evolution, while the system has not proceeded to a state where a role of
dissipation is significant due to large gradients. Besides this, 
conservation laws in physical systems, as a rule, are associated with
definite geometrical objects. Usage of these associations promotes 
understanding and vivid imagination of everything that happens. In
hydrodynamic models, the frozen-in vortex lines are such geometrical
objects, so the present work is devoted to study of motion of vortex lines
in spatially inhomogeneous systems.

Hydrodynamic equations  describe, in particular, an interaction
between "soft" degrees of freedom of a system  -- frozen-in vortices,
and "hard" degrees of freedom -- acoustic modes. The presence of soft 
degrees of freedom is explained by the fact that equilibrium states 
of the fluid are highly degenerated due to the relabeling symmetry.
Thus, no potential energy corresponds to soft degrees of freedom, unlike
the hard degrees of freedom. Due to dominating effect
of elastic potential energy, hard degrees of freedom behave, typically,
like a set of weakly nonlinear oscillators. On the contrary, dynamics
of soft degrees of freedom is not dominated by a potential energy, and 
usually it is highly nonlinear. In a limit of slow flows, 
when a typical velocity of vortex structure motion is small in comparison 
with the sound speed, a dynamic regime is possible, in which the hard 
degrees of freedom, corresponding to deviations of fluid density 
$\rho(\g{r},t)$ from an equilibrium configuration $\rho_0(\g{r})$, are 
excited weakly. Then, completely neglecting the sound, in the 
homogeneous case $\rho_0=const$ one arrives at the models of 
incompressible fluid.
For dynamics of vortices in incompressible perfect fluid the so called
formalism of vortex lines has been developed recently
\cite{Berdichevsky97}-\cite{KR2000}, which takes into account the 
conservation of topology of the vorticity field \cite{MonSas},\cite{RA94}. 
Application
of this formalism allows one to deal with a partially integrated
system, where the topology is fixed by the Cauchy invariant \cite{Lamb}.
In proposed description the frozen-in solenoidal vorticity field is considered 
as a continuous distribution of the elementary objects  -- vortex lines. 
Such formulation of inviscid hydrodynamics as the problem of vortex line
motion has been very suitable for study of localized vortex structures like
vortex filaments. Also, it seems to be an adequate approach to the problem of
finite time singularity formation in solutions of hydrodynamic equations
\cite{RPR2001PRE}.

The goal of the present work is to extend the vortex line formalism to the case 
when equilibrium density $\rho_0(\g{r})$ is a fixed nontrivial function of
spatial coordinates due to a static influence of some external forces.
Such situation takes place in many physically important models.
For examples, it can be the gravitational force for a large mass of an
isentropic gas, both in usual and in relativistic hydrodynamics, or it can
be the condition of electrical neutrality for the electron fluid on a
given background of ion distribution in the model of electron
magnetohydrodynamics (EMHD). The theory developed can be also 
applied to tasks about long-scale dynamics of the quantized vortex filaments
in a Bose-Einstein condensate placed into a trap of a sufficiently large
size. The vortex line formalism seems to be a universal and
adequate tool for investigation of slow inviscid flows in inhomogeneous 
systems. For instance, it makes possible, in a simple and standard 
way, to analyze qualitative behavior of vortices without detailed 
consideration of basic equations of motion for the fluid. Therefore the 
proposed approach can have advantages over other methods when complicated 
systems will be studied.

As a concrete result, the local induction approximation (LIA) in vortex 
dynamics will be analyzed for several spatially
inhomogeneous physical systems, namely for Eulerian compressible 
hydrodynamics in an external field, for EMHD, and for vortices in trapped 
Bose-Einstein condensates. 
A new equation of vortex filament motion will be derived, 
which takes into account the inhomogeneity of these systems 
[the Eq.(\ref{LIA})].  
As to relativistic hydrodynamics in a static
gravitational field, the proposed method gives a more complicated LIA equation
than the Eq.(\ref{LIA}), as it has been shown recently by the
present author  \cite{gr-qc/0008002}. [See also the most recent paper 
\cite{RP2001} about dynamics of an ultrarelativistic fluid in the flat 
anisotropic cosmological models of expanding Universe, 
where the formalism of vortex lines has been applied to systems with 
Hamiltonian functionals depending explicitly on the time variable, 
and the effect of nonstationary anisotropy of the space on vortex dynamics 
has been studied.]

This paper is organized as follows. A short review of Lagrangian 
formalism for fluid media is given in Sec. II. It provides a basis for
development in Sec. III of the vortex line formalism for spatially 
inhomogeneous systems. Then in Sec. IV, the method developed is applied 
to derive approximate equations of motion for slender nonstretched 
vortex filaments in three above mentioned physical models.

\section{Lagrangian formalism for a fluid}

{} From viewpoint of the Lagrangian formalism, the freezing-in 
property of the canonical vorticity is due to the special symmetry of the 
basic equations of ideal hydrodynamics \cite{ZK97}-\cite{Berdichevsky97},
\cite{R99},\cite{KR2000}. 
As known, the entire Lagrangian description of a motion of some continuous
medium can be given by the three-dimensional (3D) 
mapping $\g{r}=\g{x}(\g{a},t)$, which
indicates the space coordinates of each medium point labeled by a label
$\g{a}=(a_1,a_2,a_3)$, at an arbitrary moment in time $t$. The labeling
$\g{a}$ can be chosen in such a manner that the amount of matter in a
small volume $d^3\g{a}$ in the label space is simply equal to this volume.
With neglecting all dissipative processes, a dynamic model describing flow 
is conservative, so the equations of motion for the mapping 
$\g{x}(\g{a},t)$ follow from a variational principle
$$
\delta S=\delta\int{\cal L}\{\g{x}(\g{a},t),\dot\g{x}(\g{a},t)\}dt=0,
$$
where the Lagrangian ${\cal L}$ is a functional of 
$\g{x}(\g{a},t),\,\dot\g{x}(\g{a},t)$, and 
also spatial derivatives. A very important
circumstance is related to the fluidity property of the media under
consideration. The fluidity is manifested in the fact that the Lagrangian
actually contains the dependence on $\g{x}(\g{a},t)$ and 
$\dot\g{x}(\g{a},t)$ only through two
Eulerian characteristics of the flow, namely through the field of
density $\rho(\g{r},t)$ and the velocity field $\g{v}(\g{r},t)$, i.e.,
${\cal L}={\cal L}\{\rho,\g{v}\}$, with
\begin{equation}\label{rho_v_def}
\rho(\g{r},t)={\mbox{det}\Big\|
\frac{\partial\g{a}(\g{r},t)}{\partial\g{r}}\Big\|},
\quad
\g{v}(\g{r},t)=\dot{\g{x}}(\g{a},t)|_{\g{a}=\g{a}(\g{r},t)}
\end{equation}
Here $\g{a}(\g{r},t)$ is the inverse mapping with respect to 
${\bf x}({\bf a},t)$.

A simple particular example is the Lagrangian of ordinary Eulerian
isentropic hydrodynamics
\begin{equation}
{\cal L}_{Euler}=
\int\left(\rho\frac{\g{v}^2}{2}-\varepsilon(\rho)-
\rho U(\g{r})\right)d\g{r},
\end{equation} 
where $\varepsilon(\rho)$ is the internal energy density, $U(\g{r})$ is 
the external force potential, for instance, the gravitational potential.

A less trivial example is the Lagrangian of relativistic isentropic
hydrodynamics \cite{gr-qc/0008002} in a curved space-time with metric 
tensor $g_{ik}(t,{\bf r})$ ($i,k=0..3;\,\alpha,\beta=1..3$)
$$
{\cal L}_r=-\int{\cal E}\left(\frac{\rho}{\sqrt{-g}}
\sqrt{g_{00}+2g_{0\alpha}v^\alpha+
g_{\alpha\beta}v^\alpha v^\beta}\right)\sqrt{-g}d\g{r}.
$$
Here $g=\mbox{det}\|g_{ik}\|$ is the determinant of the metric tensor,
the expression in parenthesis is equal to the absolute 
value of the current four-vector $n^i=n(dx^i/ds)$ \cite{LL2}.
A dependence ${\cal E}(n)$ connects the relativistic density ${\cal E}$ 
of the fluid energy, measured in a locally co-moving reference frame,
with $n$.

In plasma physics, the model of electron magnetohydrodynamics is useful.
EMHD follows in the limit of slow flows from the Lagrangian of electron
fluid
\begin{equation} \label{L_e}
{\cal L}_{e}\!=\!\int\!\left(\rho\frac{\g{v}^2}{2}+
\frac{e}{mc}\rho({\bf A}\cdot{\bf v})
-\frac{(\mbox{curl\,}{\bf A})^2}{8\pi}+\dots\right)d\g{r},
\end{equation}
where $\rho({\bf r},t)$ is the density of electron fluid, 
$e$ is the electric charge of electron, $m$ is its mass, and $c$ is the
speed of light. The vector potential ${\bf A}({\bf r},t)$ of the 
electromagnetic field determines the magnetic field ${\bf B}({\bf r},t)$ 
by the relation ${\bf B}=\mbox{curl\,}{\bf A}$. In this paper, we will not 
need an explicit form of other terms indicated by the dots.

The list of examples, of course, is not exhausted by three given models.
All known hydrodynamic models without dissipation, 
where the conservation of fluid amount 
takes place, can be described in this way. So the theory developed
here is quite universal and applicable in various branches of physics
where vortex phenomena occur.

It follows from the definitions (\ref{rho_v_def}) that dynamics of the 
density $\rho(\g{r},t)$ obeys the continuity equation in its standard form
\begin{equation} \label{rho_t}
\rho_t+\nabla(\rho{\bf v})=0.
\end{equation}

The vanishing condition for variation of the action
$S=\int{\cal L}\{\rho,\g{v}\}dt$, when the mapping $\g{x}(\g{a},t)$
is varied by $\delta\g{x}(\g{a},t)$, 
can be expressed in Eulerian representation as follows (the 
generalized Euler equation \cite{R99})
\begin{equation} \label{dynequation} 
(\partial_t+{\bf v\cdot\nabla})
\left(\frac{1}{\rho}\cdot\frac{\delta {\cal L}}{\delta {\bf v}}\right)=
\nabla\left(\frac{\delta {\cal L}}{\delta \rho}\right)-
\frac{1}{\rho}\left(\frac{\delta {\cal L}}{\delta v^\alpha}\right)
\nabla v^\alpha.
\end{equation}
This is merely the variational Euler-Lagrange equation 
$$
\frac{d}{dt}\frac{\delta{\cal L}}{\delta\dot\g{x}(\g{a})}=
\frac{\delta{\cal L}}{\delta\g{x}(\g{a})}
$$
for fluid particle dynamics.
The equations (\ref{rho_t}) and (\ref{dynequation}) determine completely
evolution of hydrodynamic system.

It was already mentioned that in all such systems an
infinite number of conservation laws exists. 
The $\{\rho,{\bf v}\}$-dependence means that the Lagrangian 
${\cal L}\{\g{x}(\g{a},t),\dot\g{x}(\g{a},t)\}$ 
admits the infinite-parametric
symmetry group -- it assumes the same value on any two mappings 
$\g{x}_1(\g{a},t)$ and $\g{x}_2(\g{a},t)$, if they differ one from another
only by some relabeling of the labels with unit Jacobian
\begin{equation} \label{relabeling}
\g{x}_2(\g{a},t)=\g{x}_1(\g{a}^*(\g{a}),t),
\qquad
\mbox{det}\|\partial\g{a}^*/\partial\g{a}\|=1.
\end{equation}
Obviously, such mappings create the same density and velocity fields.
According to the Noether's theorem \cite{Arnold},\cite{DNF},  every
one-parametric sub-group of the relabeling group $\g{a}^*(\g{a})$ with
unit Jacobian corresponds to an integral of motion. There are several
classifications of these conservation laws. For instance, one can postulate
that circulation of the canonical momentum $\g{p}(\g{r},t)$,
\begin{equation} \label{p_def}
\g{p}(\g{r},t)\equiv\frac{\delta{\cal L}}{\delta\dot\g{x}(\g{a},t)}
\Big|_{\g{a}=\g{a}(\g{r},t)}
=\frac{1}{\rho}\left(\frac{\delta {\cal L}\{\rho,\g{v}\}}
{\delta {\bf v}}\right),
\end{equation}
along an arbitrary frozen-in closed contour $\gamma(t)$ does not depend on
time (the generalized theorem of Kelvin):
$$
\oint_{\gamma(t)}(\g{p}\cdot d\g{r})=const.
$$
We arrive at a different formulation, in terms of the so called Cauchy 
invariant, when consider the solenoidal field of the canonical vorticity 
$\g\Omega(\g{r},t)$,
\begin{equation} \label{Omega_def}
\g{\Omega}(\g{r},t)=\mbox{curl}\,\g{p}(\g{r},t).
\end{equation}
It is easy to check that application of the curl-operator to the equation
(\ref{dynequation}) gives
\begin{equation} \label{Omega_motion}
\g{\Omega}_t=\mbox{curl}[\g{v}\times\g{\Omega}].
\end{equation}
The formal solution of this equation is
\begin{equation} \label{Om_r}
{\bf\Omega}({\bf r},t)
=\int\delta({\bf r}-{\bf x}({\bf a},t))
({\bf\Omega}_0({\bf a})\nabla_{\bf a}){\bf x}({\bf a},t)d{\bf a},
\end{equation}
where the solenoidal independent 
on time field ${\bf\Omega}_0({\bf a})$ is exactly the Cauchy invariant. 
The equation (\ref{Om_r}) displays 
that lines of initial solenoidal field ${\bf\Omega}_0({\bf a})$
are deformed in course of motion by the mapping ${\bf x}({\bf a},t)$, 
keeping all the topological characteristics unchanged 
\cite{MonSas},\cite{RA94}. 
This feature of vortex lines dynamics is called the freezing-in property.

\section{Hamiltonian dynamics of vortex lines}

To continue, it is more convenient to reformulate 
the problem in terms of density and canonical momentum. 
Let the system be specified by some Hamilton's
functional ${\cal H}\{\rho,\g{p}\}$
\begin{equation}\label{Hdef}
{\cal H}=
\int \Big(\frac{\delta{\cal L}}{\delta {\bf v}}
\cdot{\bf v}\Big)d{\bf r}-{\cal L},
\end{equation}
where the velocity $\g{v}$ is expressed through the momentum $\g{p}$ and
through the density $\rho$ with the help of Eq. (\ref{p_def}). 
Let us note that the following equality takes place
\begin{equation}\label{vdef}
{\bf v}=\frac{1}{\rho}\left(\frac{\delta {\cal H}}{\delta {\bf p}}\right),
\end{equation}
which is analogous to the formula (\ref{p_def}). The Hamiltonian
(non-canonical \cite{ZK97}) equations of motion for the fields of density and 
momentum follow from Eqs. (\ref{rho_t}) and (\ref{dynequation}). With taking into
account the equality (\ref{vdef}), they have the form (for detailed
derivation see \cite{R99})
\begin{equation}\label{drho/dt}
\rho_t+\nabla\left(\frac{\delta {\cal H}}{\delta {\bf p}}\right)=0,
\end{equation}
\begin{equation}\label{dp/dt}
{\bf p}_t=\left[\left(\frac{\delta {\cal H}}{\delta {\bf p}}\right)
\times\frac{\mbox{curl}\ {\bf p}}{\rho}\right]
-\nabla\left(\frac{\delta {\cal H}}{\delta \rho}\right).
\end{equation}

It is supposed in this work that the Hamiltonian has a minimum at some
configuration $\{\rho_0(\g{r}),\g{p}_0(\g{r})\}$. For simplicity, we will
consider only systems without gyroscopic effects, i.e., 
$\g{p}_0(\g{r})=\g{0}$. Our purpose is to study slow flows near the 
equilibrium.
In the regime under consideration, which corresponds formally to
"prohibition" of excitation of the acoustic modes, the flow of fluid
occurs in such a way that the density of each moving portion of fluid
follows the given function $\rho_0(\g{r})$.
Therefore the equation (\ref{drho/dt}) gives the condition
\begin{equation}\label{density}
\nabla\left(\frac{\delta {\cal H}}{\delta {\bf p}}\right)=0,
\end{equation}
which means that after imposing the constrain $\rho=\rho_0(\g{r})$ the
Hamiltonian does not depend anymore on the potential component of the
canonical momentum field, it depends now only on the solenoidal component,
i.e., actually on the vorticity $\g{\Omega}$. It should be also noted that
it is sufficient to take into consideration only quadratic part of the
Hamiltonian, because the flow is supposed to be 
slow, so higher order terms, if any exists, may be neglected. Therefore, in
further equations, ${\cal H}\{\g{\Omega}\}$ is actually a quadratic
functional of the vorticity field, though this fact will not be used in 
formal calculations.
The condition (\ref{density}) implies validity of the formula
\begin{equation}
\frac{\delta {\cal H}}{\delta {\bf p}}=\mbox{curl}
\left(\frac{\delta {\cal H}}{\delta {\bf \Omega}}\right),
\end{equation}
so the next equation for slow dynamics of the vorticity follows from 
(\ref{dp/dt})
\begin{equation}\label{Ham}
{\bf\Omega}_t=\mbox{curl}
\left[\mbox{curl}\left(\frac{\delta{\cal H}}{\delta{\bf\Omega}}\right)
\times\frac{{\bf\Omega}}{\rho_0(\g{r})}
\right]
\end{equation}
This equation differs only by presence of the function $\rho_0(\g{r})$ 
(instead of the unity) from the equation used in the works 
\cite{KR98} and \cite{KR2000} as a start point in the transition to the
vortex line representation in homogeneous systems. Therefore all the
further constructions will be done similarly to Ref.\cite{KR2000}. First,
let us fix the topology of the vorticity field by means of the formula
$$
{\bf \Omega }({\bf r},t)=\int \delta({\bf r}-{\bf R}({\bf a},t))
({\bf \Omega}_{0}({\bf a})\nabla _{{\bf a}}){\bf R}({\bf a},t)d{\bf a}
$$
\begin{equation}\label{OmegaR}
=\frac{({\bf \Omega}_{0}({\bf a})\nabla _{{\bf a}}){\bf R}({\bf a},t)}
{\mbox{det}\|\partial{\bf R}/\partial{\bf a}\|}
\Big|_{{\bf a}={\bf R}^{-1}({\bf r},t)}
\end{equation}
where ${\bf \Omega}_{0}({\bf a})$ is the Cauchy invariant. The vector
\begin{equation}\label{T}
{\bf T}({\bf a},t)=
({\bf \Omega}_{0}({\bf a})\nabla _{{\bf a}}){\bf R}({\bf a},t)
\end{equation}
is directed along the vorticity field at the point
${\bf r}={\bf R}({\bf a},t)$. It is necessary to stress that the
information supplied by the mapping ${\bf R}({\bf a},t)$ is not so full as
the information supplied by the purely Lagrangian mapping 
${\bf x}({\bf a},t)$. The role of the mapping ${\bf R}({\bf a},t)$ is
exhausted by a continuous deformation of the vortex lines of the initial
field ${\bf \Omega}_{0}$. This means that the Jacobian
\begin{equation}\label{Jacobian}
J=\mbox{det}\|\partial{\bf R}/\partial{\bf a}\|
\end{equation}
is not related directly to the density $\rho_0(\g{r})$, inasmuch as, 
unlike the mapping $\g{x}(\g{a},t)$, the new mapping 
${\bf R}({\bf a},t)$  is defined up to an arbitrary non-uniform shift
along the vortex lines. Geometrical meaning of the representation
(\ref{OmegaR}) becomes more clear if instead of $\g{a}$ we use a so
called vortex line coordinate system 
$(\nu_1(\g{a}),\nu_2(\g{a}),\xi(\g{a}))$, so that the 2D Lagrangian coordinate
$\nu=(\nu_1,\nu_2)\in{\cal N}$ is a label of vortex lines, which lies in
some manifold  ${\cal N}$, while a longitudinal coordinate $\xi$
parameterizes the vortex line. Locally, vortex line coordinate system
exists for arbitrary topology of the vorticity field, but globally -- only
in the case when all the lines are closed. In the last case the equation 
(\ref{OmegaR}) can be rewritten in the simple form 
\begin{equation}\label{lines}
{\bf \Omega }({\bf r},t)=\int_{\cal N}d^2\nu \oint \delta ({\bf r}-
{\bf R}(\nu,\xi,t)){\bf R}_{\xi}d\xi,  
\end{equation}
where ${\bf R}_{\xi}=\partial {\bf R}/\partial\xi$.
The geometrical meaning of this formula is rather evident --- the frozen-in
vorticity field is presented as a continuous distribution of vortex
lines. It is also clear that the choice of the longitudinal parameter is
nonunique. This choice is determined exclusively by convenience for
a particular task.
Usage of  the formula
\begin{equation}\label{OmegaRt}
{\bf\Omega}_t({\bf r}\!,t)\!=\!
\mbox{curl}_{\bf r}\!\int\!
\delta({\bf r}\!-\!{\bf R}({\bf a},\!t))
[{\bf R}_t({\bf a},t)\!\times\!{\bf T}({\bf a},t)]
d{\bf a},
\end{equation}
which follows immediately from Eq. (\ref{OmegaR}), together with the
general relationship between variational derivatives of an arbitrary
functional $F\{\g{\Omega}\}$
\begin{equation}\label{peresch}
\left[ {\bf T}\times \mbox{curl}_{\bf r}\left(\frac{\delta F}
{\delta{\bf \Omega}
({\bf R})}\right) \right] =
\frac{\delta F\{{\bf \Omega }\{{\bf R}\}\}}{\delta {\bf R}({\bf a})}
\Big|_{{\bf \Omega }_{0}}  
\end{equation}
allow us to obtain the equation of motion for the mapping 
${\bf R}({\bf a},t)$ by substitution of the representation (\ref{OmegaR}) 
into the equation (\ref{Ham}). As the result, dynamics of the mapping 
${\bf R}({\bf a},t)$ is determined by the equation
\begin{equation}\label{main}
\left[ ({\bf \Omega }_{0}({\bf a})\nabla _{{\bf a}}){\bf R}({\bf a})\times
{\bf R}_{t}({\bf a})\right]\rho_0(\g{R}) =
\frac{\delta {\cal H}\{{\bf \Omega }\{{\bf R}\}\}}
{\delta {\bf R}({\bf a})}.  
\end{equation}
It is not very difficult to check by a direct calculation that the given
equation of motion for ${\bf R}({\bf a},t)$ 
follows from the variational principle 
$\delta\int{\cal L}_{{\bf \Omega}_{0}}dt=0$, where the Lagrangian is 
\begin{equation}
{\cal L}_{{\bf \Omega}_{0}}\!=\!
\int\!\Big(\!\left[ {\bf R}_{t}\times
{\bf D}({\bf R})\right]
\!\cdot\!({\bf \Omega }_{0}\nabla _{{\bf a}})
{\bf R}\Big)d{\bf a}
 - {\cal H}\{{\bf \Omega }\{{\bf R}\}\!\},  \label{LAGRANGIAN}
\end{equation}
with the vector function ${\bf D}({\bf R})$ being related to the
density $\rho_0(\g{r})$ by the equality
\begin{equation}\label{divD}
(\nabla_{\bf R}\cdot{\bf D}({\bf R}))=\rho_0(\g{R}).
\end{equation}
For application to vortex filaments, the following form of the
Lagrangian is more useful, where ${\bf R}={\bf R}(\nu,\xi,t)$:
\begin{equation}
{\cal L}_{\cal N}=\!
\int_{\cal N}\!\!d^2\nu\!\oint\! \Big(\left[ {\bf R}_{t}\times
{\bf D}({\bf R})\right]
\!\cdot\!
{\bf R}_\xi\Big)d\xi 
 - {\cal H}\{{\bf \Omega }\{{\bf R}\}\!\}.  \label{LAGR_lines}
\end{equation}

It should be stressed that conservation in time of the fluid amount inside
each closed frozen-in vortex surface is not imposed {\it a priori} as a
constrain for the mapping ${\bf R}({\bf a},t)$. All such quantities are
conserved in the dynamical sense due to the symmetry of the Lagrangian 
(\ref{LAGR_lines}) with respect to the group of relabelings of the labels
$\nu$ of vortex lines
\begin{equation}\label{nu_relabl}
\nu=\nu(\tilde\nu,t),\qquad
{\partial(\nu_1,\nu_2)}/
{\partial(\tilde\nu_1,\tilde\nu_2)}=1.
\end{equation}
Considering all one-parametrical subgroups of the given group of
area-preserving transformations and applying the Noether's theorem 
\cite{DNF} to the Lagrangian (\ref{LAGR_lines}),
it is possible to obtain the indicated integrals of motion in the next
form (compare with Ref.\cite{R2000})
\begin{equation}\label {IPsi}
I_{\Psi}= \int_{\cal N}\Psi(\nu_1,\nu_2) d^2\nu \oint \rho_0(\g{R})
([{\bf R}_1\times{\bf R}_2]\cdot{\bf R}_\xi)d\xi
\end{equation} 
where $\Psi(\nu_1,\nu_2)$ is an arbitrary function on the manifold 
${\cal N}$ of labels, with the only condition 
$\Psi|_{\partial{\cal N}}=0$.

\section{Local induction approximation}

When a particular task is being solved, the necessity always arises in making 
some simplifications. The variational formulation for the dynamics of vortex
lines allows us to introduce and control various approximations on the level
of the Lagrangian (\ref{LAGR_lines}), what in practice is more convenient
and more simple than control of approximations made on the level 
of equations of motion. For example we will consider now the so called
local induction approximation (LIA) in dynamics of a slender non-stretched
vortex filament. As known, in spatially homogeneous systems the LIA yields
an integrable equation which is gauge equivalent to the Nonlinear Schroedinger
Equation \cite{Hasimoto}. In general case, inhomogeneity 
destroys the integrability of LIA equation. Nevertheless, this does not
reduce the value of LIA as a simplified model of filament dynamics.

\subsection{LIA in Eulerian hydrodynamics}

At first, we will consider the Eulerian hydrodynamics,
where the canonical momentum and the velocity coincide. Let the vorticity
be concentrated in a quasi-one-dimensional structure, a vortex filament,
with a typical  longitudinal scale $L$ being much larger than the width 
$d$ of the filament. A typical scale of spatial inhomogeneity is
supposed to be of order of $L$ or larger. In such situation, the kinetic
energy of the fluid is concentrated in the vicinity of the filament, with
the corresponding integral being logarithmically large on the parameter
$L/d$. The LIA consist in the following simplifications. First, in the
kinetic part of the Lagrangian (\ref{LAGR_lines}), the dependence of
the shape of vortex lines  on the label  $\nu$ is neglected, i.e., the
filament is considered as a single curve $\g{R}(\xi,t)$. After integration
over $d^2\nu$ the constant multiplier $\Gamma$ appears now, which is the
value of velocity circulation around the filament. Second, some significant
simplifications may be done in the Hamiltonian. Generally speaking, exact
expression for the Hamiltonian implies derivation of the dependence 
$\g{v}\{\rho_0,\g{\Omega}\}$ from the following system of equations:
$$
\mbox{curl\,}\g{v}=\g{\Omega},
\quad
\mbox{div\,}(\rho_0(\g{r})\cdot\g{v})=0,
$$  
and subsequent substitution of $\g{v}$ into the expression for the
kinetic energy. After that one has to deal with a nonlocal Hamiltonian
$$
{\cal H}^{\{\rho_0\}}_{Euler}=\frac{1}{2}\int\!\int 
G^{\{E,\rho_0\}}_{\alpha\beta}(\g{r}_1,\g{r}_2)
\Omega_\alpha(\g{r}_1)\Omega_\beta(\g{r}_2)d\g{r}_1d\g{r}_2,
$$
where the Green function $G^{\{E,\rho_0\}}_{\alpha\beta}(\g{r}_1,\g{r}_2)$ 
has the following asymptotics at close arguments:
$$
G^{\{E,\rho_0\}}_{\alpha\beta}(\g{r}_1,\g{r}_2)\to 
\frac{\rho_0(\g{r}_1)\delta_{\alpha\beta}}{4\pi|\g{r}_2-\g{r}_1|}, 
\qquad \g{r}_2\to\g{r}_1.
$$
Therefore the Hamiltonian of a singular vortex filament,
\begin{equation}\label{exact_ham}
{\cal H}^{d=0}_f=\frac{\Gamma^2}{2}\oint\!\oint 
G^{\{E,\rho_0\}}_{\alpha\beta}(\g{R}_1,\g{R}_2)R_{1\alpha}'
R_{2\beta}'d\xi_1 d\xi_2,
\end{equation}
where $R_{1\alpha}'=\partial_{\xi_1} R_{\alpha}(\xi_1)$ and so on,
logarithmically diverges. Taking into account the finite width $d$ and the
longitudinal scale $L$,
it is possible to put, with a logarithmic accuracy, the Hamiltonian of
a thin vortex filament  equal to the following expression
\begin{equation}\label{LIA_ham}
{\cal H}^d_f\approx{\cal H}_A=\Gamma A\oint\rho_0(\g{R})|\g{R}_\xi|d\xi, 
\end{equation}
where the constant $A$ is
\begin{equation}
A=\frac{\Gamma}{4\pi}\ln\left(\frac L d\right).
\end{equation}
In accordance with the simplifications made above, the motion of a
slender vortex filament in the spatially inhomogeneous system is 
described approximately by the equation
$$
[\g{R}_\xi\times\g{R}_t]\rho_0(\g{R})/A=\nabla\rho_0(\g{R})\cdot |\g{R}_\xi|
-\partial_\xi\left(\rho_0(\g{R})\cdot\frac{\g{R}_\xi}{|\g{R}_\xi|}\right),
$$
which is obtained by substitution of the Hamiltonian (\ref{LIA_ham})
into the equation (\ref{main}). The given equation can be solved with
respect to $\g{R}_t$ and rewritten in terms of the geometrically invariant
objects $\g{t},\g{b},\kappa$, where $\g{t}$ is the unit tangent vector on
the curve, $\g{b}$ is the unit binormal vector, and $\kappa$ is the
curvature of the line. As the result, we have the equation
\begin{equation}\label{LIA}
\g{R}_t/A=[\nabla(\ln\rho_0(\g{R}))\times\g{t}]+\kappa\g{b},
\end{equation}  
applicability of which is not limited actually by the Eulerian
hydrodynamics. Let us indicate at least two more physical models where the
LIA equation (\ref{LIA}) is useful.

\subsection{LIA in EMHD}

The first model is EMHD, the Hamiltonian of which contains, besides 
the kinetic energy, also the energy of 
magnetic field $\g{B}$ created by current of the electron fluid
through the motionless inhomogeneous ion fluid. In principle, the
Hamiltonian of EMHD is determined by the relations that follow from the
Lagrangian ${\cal L}_e$, Eq.(\ref{L_e}):
$$
\mbox{curl\,}\g{v}+\frac{e}{m c}\g{B}=\g{\Omega},\qquad
\mbox{curl\,}\g{B}=\frac{4\pi e}{m c}\rho_0(\g{r})\cdot\g{v},
$$
\[
{\cal H}_{EMHD}=\int\left(\rho_0(\g{r})\frac{\g{v}^2}{2}+
\frac{\g{B}^2}{8\pi}\right)d\g{r}.
\]
In spatially homogeneous system we would obtain the expression 
$$
{\cal H}^{h}_{EMHD}=\frac{\rho_0}{8\pi}\int\!\!\int
\frac{e^{-q|{\bf r}_1-{\bf r}_2|}}{|{\bf r}_1-{\bf r}_2|}
{\bf\Omega}({\bf r}_1)\cdot{\bf\Omega}({\bf r}_2)d{\bf r}_1d{\bf r}_2,
$$
where the screening parameter $q$ is determined by the relation
$$
q^2=\frac{4\pi\rho_0e^2}{m^2c^2}.
$$
In inhomogeneous system $q$ is a function of coordinates, with a
typical value $\tilde q$. Let us suppose the inequalities $\tilde q L\gg1$,
and $\tilde q d\ll1$.
One can see that the logarithmic integral analogous to the expression
(\ref{exact_ham}) is cut now
not on the $L$, but on the skin depth $\lambda=1/q$. Accordingly, for this
case the constant $A$ in LIA equation  (\ref{LIA}) is given by the
expression
$$
A_{EMHD}=
\frac{\Gamma}{4\pi}\ln\left(\frac {mc} {ed\sqrt{\tilde\rho}}\right).
$$
We see that in ideal EMHD the LIA works better than in Eulerian
hydrodynamics, due to the screening effect.

\subsection{LIA in Bose-Einstein condensate}

Another important physical model, where the equation (\ref{LIA}) may 
be applied, is the theory of Bose-Einstein condensate for a weakly 
nonideal trapped gas with a quantized vortex filament \cite{BEcond}.
At zero temperature this system is described
approximately by the complex order parameter $\Phi(\g{r},t)$ (the wave 
function of the condensate), with the equation of motion (the
Gross-Pitaevskii equation) taking in dimensionless variables the form
\begin{equation}\label{GP}
i\Phi_t=\left(-\frac{1}{2}\Delta+U(\g{r})-\mu+|\Phi|^2\right)\Phi,
\end{equation}
where $U(\g{r})$ is an external potential, usually of the quadratic form
$$
U(\g{r}) =ax^2+by^2+cz^2,
$$ 
and the constant $\mu$ is the chemical potential. Let us suppose 
$a\ge b\ge c$. It is well known that the equation (\ref{GP}) admits the
hydrodynamical interpretation. The variables $\rho$ and $\g{p}$ are
defined by the relations
$$
\rho=|\Phi|^2,\qquad
\rho\g{p}=(\bar\Phi\nabla\Phi-\Phi\nabla\bar\Phi)/2i.
$$
The corresponding Hamiltonian is 
$$
{\cal H}_{GP}=\int\left[
\frac{(\nabla\sqrt{\rho})^2+\rho\g{p}^2}{2}+(U(\g{r})-\mu)\rho
+\frac{\rho^2}{2}\right]d\g{r}.
$$
In comparison with the ordinary Eulerian hydrodynamics, there is the term
depending on the density gradient in this expression. However, with large
values of the parameter $\mu^2/a$, one may neglect that term in
calculation of the equilibrium density inside the space region where the
density is not exponentially small, and use the approximate formula
$$
\rho_0(\g{r})\approx\mu-U(\g{r}),\qquad \mbox{if}
\quad \{\mu^2\gg a; \quad \mu-U(\g{r})>0\}.
$$

As known, the equation (\ref{GP})  admits solutions with
quantized vortex filaments, the circulation around them being equal to
$2\pi$. In these solutions, the density differs significantly from 
$\rho_0(\g{r})$ only at close distances of order $1/\sqrt{\mu}$ from the
zero line. Far away, up to distances of order 
$L\sim\sqrt{\mu/a}\gg 1/\sqrt{\mu}$, we have almost Eulerian flow. 
Therefore the LIA 
equation (\ref{LIA}) is valid for description of slow motion of quantum 
vortex filament in trapped Bose-condensate of a relatively large size
$L$, with the parameter 
$$
A=A_{GP}=({1}/{4})\ln({\mu^2}/{a}).
$$
The inequality $\mu^2\gg a$ ensures also the smallness of the filament
velocity $v_f\sim \kappa A_{GP}$ with respect to the speed of sound 
$c_s\sim\sqrt{\mu}$,  while the curvature of the filament is of order 
$\kappa\sim\sqrt{a/\mu}$ .

\section*{Conclusions}

Let us summarize briefly the main results of this paper. First, with 
neglecting acoustic degrees of freedom in investigation of slow isentropic 
flows of a compressible perfect fluid in spatially inhomogeneous systems,  
the general form of variational principle for dynamics of frozen-in 
vortex lines has been found. The connection from the basic Lagrangian 
given in terms of density and velocity fields to the Hamiltonian of vortex 
lines has been provided, which allows one to analyze vorticity dynamics 
in complicated systems initially specified by the principle of least action.
Second, this method has been applied to several physically important
models, such as Eulerian hydrodynamics in an external field, ideal 
electron magnetohydrodynamics on inhomogeneous ion background, 
and the Gross-Pitaevskii model for trapped Bose-Einstein condensate, 
in order to derive approximate equations
of motion for vortex filaments. It has been established that a 
mathematical structure of the equations derived is the same in all the three 
cases, though parameters have different physical meaning in each case.

The final remark concerns possibility for development of an analogous approach
in the general case, when acoustic waves are important. Though at present
moment this work has not been completed yet, but the author hopes it will be
done in the future on the basis of equations (\ref{drho/dt}-\ref{dp/dt}).

\section*{Acknowledgments}

This work was supported by RFBR (grant No. 00-01-00929), by the Russian State
Program of Support of the Leading Scientific Schools (grant No. 00-15-96007), 
by the INTAS, and by the Fund KFA, Forschungszentrum, Juelich, Germany.

\end{document}